\documentclass[preprint, 11pt]{elsarticle}

\usepackage{amssymb}
\usepackage{verbatim}
\usepackage{color}
\usepackage{amsthm}
\usepackage{amsmath}
\usepackage[dvipsnames]{xcolor}
\usepackage{tikz}
\usepackage{amsxtra}
\usepackage{caption}
\usepackage{subcaption}
\usepackage{algorithm,algcompatible}
\algnewcommand\INPUT{\item[\textbf{Input:}]}
\algnewcommand\OUTPUT{\item[\textbf{Output:}]}%
\usepackage{lineno}

\usepackage{graphicx}

\DeclareRobustCommand{\blackline}{\raisebox{2pt}{\tikz{\draw[Black,solid,line width = 0.9pt](0,0) -- (5mm,0);}}}
\DeclareRobustCommand{\blueline}{\raisebox{2pt}{\tikz{\draw[RoyalBlue,solid,line width = 0.9pt](0,0) -- (5mm,0);}}}
\DeclareRobustCommand{\orangeline}{\raisebox{2pt}{\tikz{\draw[BurntOrange,solid,line width = 0.9pt](0,0) -- (5mm,0);}}}
\DeclareRobustCommand{\greenline}{\raisebox{2pt}{\tikz{\draw[Green,solid,line width = 0.9pt](0,0) -- (5mm,0);}}}
\DeclareRobustCommand{\redline}{\raisebox{2pt}{\tikz{\draw[Red,solid,line width = 0.9pt](0,0) -- (5mm,0);}}}

\DeclareRobustCommand{\blackdashedline}{\raisebox{2pt}{\tikz{\draw[Black,dashed,line width = 0.9pt](0,0) -- (5mm,0);}}}
\DeclareRobustCommand{\blackdottedline}{\raisebox{2pt}{\tikz{\draw[Black,dotted,line width = 0.9pt](0,0) -- (5mm,0);}}}
\DeclareRobustCommand{\reddashedline}{\raisebox{2pt}{\tikz{\draw[Red,dashed,line width = 0.9pt](0,0) -- (5mm,0);}}}
\DeclareRobustCommand{\reddottedline}{\raisebox{2pt}{\tikz{\draw[Red,dotted,line width = 0.9pt](0,0) -- (5mm,0);}}}
\DeclareRobustCommand{\bluedashedline}{\raisebox{2pt}{\tikz{\draw[RoyalBlue,dashed,line width = 0.9pt](0,0) -- (5mm,0);}}}

\DeclareRobustCommand{\purpledottedline}{\raisebox{2pt}{\tikz{\draw[Purple,dotted,line width = 0.9pt](0,0) -- (5mm,0);}}}

\DeclareRobustCommand{\tikzcircle}[2][red,fill=red]{\tikz[baseline=-0.5ex]\draw[#1,radius=#2] (0,0) circle ;}

\usepackage{lineno}   %% <- no mathlines option
\usepackage{amsmath}  %% <- after lineno
\usepackage{etoolbox} %% <- for \cspreto, \csappto

\newcommand*\linenomathpatch[1]{%
  \cspreto{#1}{\linenomath}%
  \cspreto{#1*}{\linenomath}%
  \csappto{end#1}{\endlinenomath}%
  \csappto{end#1*}{\endlinenomath}%
}

\linenomathpatch{equation}
\linenomathpatch{gather}
\linenomathpatch{multline}
\linenomathpatch{align}
\linenomathpatch{alignat}
\linenomathpatch{flalign}

\journal{Comput Methods Appl Mech Eng}

\begin{document}

\begin{frontmatter}

\title{Sequential Bayesian experimental design for estimation of extreme-event probability in stochastic input-to-response systems}

\author{Xianliang Gong}
%\ead{xlgong@umich.edu}
\author{Yulin Pan\corref{cor1}}
\ead{yulinpan@umich.edu}
\address[a]{Department of Naval Architecture and Marine Engineering, University of Michigan, 48109, MI, USA}
\cortext[cor1]{Corresponding author}
%\affiliation[1]{organization={Department of Naval Architecture and Marine Engineering, University of Michigan}, 
%                 addressline={2600 Draper Dr},
%                 postcode={48109}, 
%                 city={Ann Arbor}, 
%                 state = {MI}
%                 country={USA}}

\begin{abstract}
%\st{Hello World}
We consider an input-to-response (ItR) system characterized by (1) parameterized input with a known probability distribution and (2) stochastic ItR function with heteroscedastic randomness. Our purpose is to efficiently quantify the extreme response probability when the ItR function is expensive to evaluate. The problem setup arises often in physics and engineering problems, with randomness in ItR coming from either intrinsic uncertainties (say, as a solution to a stochastic equation) or additional (critical) uncertainties that are not incorporated in a low-dimensional input parameter space (as a result of dimension reduction applied to the original high-dimensional input space). To reduce the required sampling numbers, we develop a sequential Bayesian experimental design method leveraging the variational heteroscedastic Gaussian process regression (VHGPR) to account for the stochastic ItR, along with a new criterion to select the next-best samples sequentially. The validity of our new method is first tested in two synthetic problems with the stochastic ItR functions defined artificially. Finally, we demonstrate the application of our method to an engineering problem of estimating the extreme ship motion probability in irregular waves, where the uncertainty in ItR naturally originates from standard wave group parameterization, which reduces the original high-dimensional wave field into a two-dimensional parameter space. 

\end{abstract}



\end{frontmatter}


%\linenumbers
\section{Introduction}

Extreme events, considered as abnormally large observables, can happen in many natural and engineering systems \cite{farazmand2019extreme,ghil2011extreme,rahmstorf2011increase}. Examples of such events include tsunami, extreme precipitation and extreme structural vibrations, which often cause catastrophic consequences to the society, industry and environment. The quantification of the probability of extreme events is of vital importance for the design of the engineering systems to alleviate their devastating impact.

Many systems of interest are characterized by an input-to-response (ItR) function which can only be queried by expensive experiments or simulations. For example, the ship motion response induced by an input wave group is obtained from expensive models of computational fluid dynamics (CFD). If the ItR function is deterministic, the probability distribution of the response is then induced only by the (assumed) known probability distribution of the input. The quantification of extreme-event probability for such systems has been studied extensively in the context of Monte-Carlo simulations. Due to the expensive evaluations of ItR and rareness of the extreme events, many studies aim for reducing the number of required samples in the computation. While techniques such as importance sampling and control variate \cite{mcbook} provide certain levels of acceleration, a more successful category of methods make use of the sequential Bayesian experimental design (BED) \cite{chaloner1995bayesian} (or active learning \citep{cohn1996active}), where the next-best samples are selected based on the existing information. Two key components of the sequential BED are a surrogate model (usually a Gaussian progress regression \citep{rasmussen2003gaussian}) to approximate the ItR, and an optimization problem maximizing a predefined acquisition function to select the next-best samples. Many sequential BED methods have been developed for the purpose of estimating extreme-event probability, with varying acquisition functions, e.g., AK-MCS \citep{echard2011ak}, EGRA\citep{bichon2008efficient}, GSAS\citep{hu2016global}, information-theoretic  design \citep{wang2016gaussian}, and output-weighted sequential sampling \citep{mohamad2018sequential, blanchard2020output}.

%Varying in details in the two components, a spectrum of sequential BED methods for the purpose of estimating extreme-event probability have been developed 

However, in many cases, the ItR has to be considered as a stochastic function, instead of a deterministic one. These situations may originate from (a) an intrinsically stochastic dynamical system, e.g. stochastic differential equations modeling a physical diffusion process or stock prices \citep{holden1996stochastic}; the stochastic model of climate variability including the non-average `weather' component as random forcing terms \citep{hasselmann1976stochastic}, and (b) some uncertain variables
that are not easily incorporated in a low-dimensional input parameter space, in particular when dimension reduction technique is applied to a high-dimensional input space resulting in uncertainties in the reduced dimensions \citep{tsilifis2021bayesian}. Under such situations, the probability distribution of the response is critically influenced by the randomness in the ItR (in addition to the probability distribution of input parameters). If the randomness of the ItR is uniform for all input parameters, previous techniques \citep{echard2011ak, bichon2008efficient, hu2016global,wang2016gaussian, mohamad2018sequential,blanchard2020output} for deterministic ItR can be extended to handle the situation (by incorporating the uniform randomness in the Gaussian process regression). However, more often, the uncertainty of the ItR is inhomogeneous for different input parameters, e.g., the uncertainty of power production of wind turbine is different among various wind speed inputs \citep{rogers2020probabilistic}. This results in a heteroscedastic ItR with the variance of response not representable as a constant. To our knowledge, currently there is no sequential BED method designed to consider heteroscedastic uncertainty in ItR, and its effect on extreme-event probability.

%To our best knowledge, the existing BED methods usually deal with deterministic ItR. A few of them introduce homogeneous randomness but treat them as useless noise. For these situations, a standard Gaussian process \cite{rasmussen2003gaussian} is qualified as a surrogate model. However, the homogeneous randomness assumption made by standard Gaussian process is not suitable for general random ItR. Meanwhile, the acquisition function should be objective (QoI) oriented and accord with the specific surrogate. The existing acquisitions \citep{mohamad2018sequential,sapsis2020output,hu2016global,echard2011ak} are not applicable to our problem. A new BED method is needed. 

In this work, we propose a new method to quantify the probability of extreme events (defined as an observable above a given threshold) considering the ItR with heteroscedastic uncertainty. The core of our algorithm is a variational heteroscedastic Gaussian process regression (VHGPR) which approximates the ItR with sufficiently low computational cost and high accuracy. This brings major improvement upon all previous BED methods employing the standard Gaussian process regression (SGPR) which are unable to resolve the heteroscedasticity in the ItR. Accordingly, we formulate a new acquisition function for selecting the next-best sample considering both the probability distribution of inputs and uncertainty in ItR. We first demonstrate the effectiveness of our method in two synthetic problems to estimate the extreme-event probability. 
We show that drastically improved performance is achieved compared to existing approaches based on SGPR. Finally, we demonstrate the superiority of our method (to existing methods) in solving an engineering problem of estimating the extreme ship motion probability in irregular waves. The difficulty in this problems lies in the heteroscedastic uncertainty of the ItR resulted from the wave group parameterization which reduces the original high-dimensional wave field to a two-dimensional parameter space. We show that the effect of this type of uncertainty to the exceeding probability can be successfully considered in our approach.

The python code package for implementation of our method and example cases in this paper, named HGPextreme, is available on Github \footnote{https://github.com/umbrellagong/HGPextreme}.  

% For random systems, the hierarchical fidelities could come from both model space (e.g. an expensive direct simulation VS an cheap analytical model) and probability space (an ).

\section{Computational framework}
\subsection{Problem setup}
We start from an ItR system with input $\boldsymbol{x} \in \mathbb{R}^d$ of known distribution $X\sim p_X(\boldsymbol{x})$ and response $y \in \mathbb{R}$. An ItR function $S$ directly relates $\boldsymbol{x}$ to $y$ with its randomness represented by $\omega$:
\begin{equation}
    y(\omega) = S(\boldsymbol{x}, \omega), \; \omega \in
    \Omega.
    \label{ItR}
\end{equation}
To be more specific, $\omega$ is a random seed lying in the sample space $\Omega$. For given $\boldsymbol{x}$, $y(\omega)$ represents a random variable, i.e., a function from sample space to real number $\Omega\rightarrow \mathbb{R}$.

Our interest is the exceeding probability of $y(\omega)$ above a threshold $\delta$:
\begin{align}
P_e \equiv \mathbb{P}(S(X,\omega)> \delta) & = \int \mathbb{P}(S(X, \omega)>\delta | X = \boldsymbol{x}) p_X(\boldsymbol{x}) d \boldsymbol{x} \nonumber \\
& = \int \mathbb{P}(S(\boldsymbol{x},\omega)>\delta) p_X(\boldsymbol{x}) d \boldsymbol{x}. \; 
\label{exprob}
\end{align}
It is clear that both distribution $p_X$ and uncertainty $\omega$ contribute to the exceeding probability in \eqref{exprob}. Moreover, the variance of the response $S$ (introduced by $\omega$) is generally different for different input $\boldsymbol{x}$, resulting in a heteroscedastic uncertainty of the ItR. We remark that this problem setup including \eqref{ItR} and \eqref{exprob} are motivated in the dicussion in Sec. 1, and resolving this heteroscedasticity in the ItR is critical for the success of our new method (or improvement of our method compared to all previous approaches) as will be discussed in Sec. 3.

A brute-force computation of \eqref{exprob} calls for extensive Monte-Carlo samples in the probability space associated with both $X$ and $\omega$, e.g. \cite{perdikaris2015multi}, which is prohibitive under expensive queries of $S(\boldsymbol{x},\omega)$. Therefore, we seek to develop a sampling algorithm following the sequential BED framework, where each sample is selected making use of the existing information of previous samples. Our new sampling algorithm also has to be developed in conjunction with the heteroscedastic uncertain ItR that has not been considered before. In summary, two key components in our new approach are (i) an inexpensive surrogate model based on the variational heteroscedastic Gaussian process regression (VHGPR) to approximate the heteroscedastic ItR; and (ii) an optimization based on an acquisition function to provide the next-best samples with fast convergence in computing \eqref{exprob}. We next describe the two components in detail in Sec. 2.2 and Sec. 2.3, followed by the overall algorithm \ref{al}.

\subsection{VHGPR as a surrogate model}
To introduce the surrogate model for the ItR, we first rewrite \eqref{ItR} as
\begin{equation}
    S(\boldsymbol{x}, \omega) = f(\boldsymbol{x}) + R(\boldsymbol{x}, \omega),
\label{ItR2}
\end{equation}
where $f(\boldsymbol{x})\equiv \mathbb{E}[S(\boldsymbol{x}, \omega)]$ is the mean of $S(\boldsymbol{x}, \omega)$ with respect to $\omega$, and $R(\boldsymbol{x}, \omega)$ is the uncertain component with zero mean. Given a dataset (from previous samples) $\mathcal{D}=\{\boldsymbol{x}^i,y^i\}_{i=1}^{i=n}$, our purpose is to approximate  \eqref{ItR2} using Gaussian process regression as involved in many sequential BED problems.

In standard Gaussian process regression (SGPR), as implemented in most previous applications for extreme-event probability, one can approximate \eqref{ItR2} as:
\begin{align}
    f(\boldsymbol{x}) | \mathcal{D} & \sim \mathcal{GP}(\mu_f(\boldsymbol{x}), {\rm{cov}}_f(\boldsymbol{x},\boldsymbol{x}')), \label{SGP1}\\
     R(\boldsymbol{x}, \omega) | \mathcal{D} & \sim \mathcal{N}(0, \gamma_0^2) \label{SGP2},
\end{align}
where $\mathcal{GP}(\cdot,\cdot)$ represents a Gaussian process with the first argument as the mean and the second argument as the covariance function. 
%Formulas of $\mu_f$ and ${\rm{cov}}_f$ are presented in \eqref{SGPPredf} and \eqref{SGPPredC}.
The uncertain component $R(\boldsymbol{x}, \omega)$ is approximated by an independent normal function at all $\boldsymbol{x}$ with constant variance $\gamma^2_0$. Clearly, the heteroscedasticity in ItR (i.e., the dependence of $R$ on $\boldsymbol{x}$) cannot be captured by the SGPR. 

% The influence of the omega is not independent. 

To incorporate the heteroscedasticity, we need to rely on the heteroscedastic Gaussian process regression (implemented as VHGPR following \cite{lazaro2011variational} in this work). In VHGPR, we are able to approximate \eqref{ItR2} as
\begin{align}
    f(\boldsymbol{x}) |\mathcal{D} & \sim \mathcal{GP}(\mu_f(\boldsymbol{x}), {\rm{cov}}_f(\boldsymbol{x},\boldsymbol{x}')), 
\label{VHGP1} \\
     R(\boldsymbol{x}, \omega) | \mathcal{D} & \sim \mathcal{N}(0, e^{g(\boldsymbol{x})}),
\label{VHGP2} \\
     g(\boldsymbol{x})|\mathcal{D}  & \sim \mathcal{GP} (\mu_g(\boldsymbol{x}), {\rm{cov}}_g(\boldsymbol{x},\boldsymbol{x}'))  
\label{VHGP3},
\end{align}
where the heteroscedastic (log) variance of the uncertain term $R(\boldsymbol{x}, \omega)$ is represented by another $\mathcal{GP}$ with mean $\mu_g(\boldsymbol{x})$ and covariance function ${\rm{cov}}_g(\boldsymbol{x},\boldsymbol{x}')$. We remark that \eqref{VHGP2} implies that the distribution associated with $\omega$ in \eqref{ItR} can be approximated by a Gaussian. Although the Gaussian assumption is a standard practice in many literature \citep{wan2017reduced, ma2021data, maulik2021latent, guo2018reduced}, we will perform a validity check of this assumption in the specific problem solved in Sec. 3.3.

Both approximations in SGPR (in terms of \eqref{SGP1}) and VHGPR (\eqref{VHGP1} and \eqref{VHGP3}) are computed as posterior predictive distributions under a Bayesian framework, with hyperparameters (say $\boldsymbol{\theta}$) determined from maximizing the likelihood function $p(\mathcal{D}|\boldsymbol{\theta})$. For SGPR, both the likelihood function and posterior \eqref{SGP1} can be derived analytically, allowing a straightforward and inexpensive numerical implementation. In contrast, for heteroscedastic GPR, the introduction of the Gaussian process on $g(\boldsymbol{x})$ prohibits analytical results on the likelihood function and posterior, posing great challenges in the numerical computation (which involves high-dimensional integration). 

In order to reduce the computational cost, variational inference is applied in VHGPR, which uses parameterized Gaussian distributions to approximate some critical distributions involved in the posterior and likelihood function. These Gaussian distributions can be determined efficiently through some optimization problems to minimize their differences from the critical distributions.
As a result of this approximation, the high-dimensional integration can be reduced to analytical formulations which leads to inexpensive computations (approximations) of the posterior and the likelihood function. In particular, the computational cost of VHGPR is only twice of SGPR, alleviating the resource requirement for the computation. More details on the algorithms of the VHGPR, along with SGPR, are summarized in Appendix A for completeness. The interested readers can also refer to \citep{rasmussen2003gaussian,lazaro2011variational} for details.
% These Gaussian distributions and the approximated likelihood can be determined efficiently through one optimization problem to minimize differences (which is cheap to compute) between proposed Gaussian distributions and the critical distributions. As a result of this approximation, we are able to obtain analytical formulations of the posterior.

In summary, the VHGPR provides us with an estimation of the ItR, $S(\boldsymbol{x}, \omega | f(\boldsymbol{x}), g(\boldsymbol{x}))$, where $f(\boldsymbol{x})|\mathcal{D}$ and $g(\boldsymbol{x})|\mathcal{D}$ follow distributions in \eqref{VHGP1} and \eqref{VHGP3} (Hereafter we will delete the condition on $\mathcal{D}$ for conciseness). Given realizations of $f(\boldsymbol{x})$ and $g(\boldsymbol{x})$, the intrinsic randomness in ItR is expected to be captured by $\omega$, i.e., the heteroscedastic distribution in \eqref{VHGP2}.

%A Gaussian (variational) distribution with unknown parameters is proposed to approximate a key distribution in computing predictions, the posterior distribution of mean and variance at training positions (latent distribution). This variational distribution is determined by minimizing its KL divergence with latent distribution, or equivalently maximizing an analytically tractable lower bound (ELBO in \eqref{ELBO}) of the likelihood. Thus we are able to obtain the optimal variational distribution and optimal hyperparamters bypass the expensive numerical or sampling-based integration. To further reduce the computational cost (number of unknown parameters in variational distribution), we assume conditional independence among mean and variance, resulting in the computational cost twice as much as SGPR. 

\subsection{Acquisition function}
Given the VHGPR of the ItR, the exceeding probability can be expressed as
\begin{equation}
\mathbb{P}(S(X, \omega | f(\boldsymbol{x}), g(\boldsymbol{x})) > \delta) = \int \mathbb{P}(S(\boldsymbol{x}, \omega | f(\boldsymbol{x}), g(\boldsymbol{x}))>\delta) p_X(\boldsymbol{x}) d \boldsymbol{x}, \label{VHGPEP}
\end{equation}
\sloppy which depends on the realizations of $f(\boldsymbol{x})$ and $g(\boldsymbol{x})$. The purpose here is to construct an acquisition function, based on which the next sample can be selected to minimize the variance of the estimation \eqref{VHGPEP}, i.e., ${\rm{var}}_{f,g}[\mathbb{P}(S(X, \omega | f(\boldsymbol{x}), g(\boldsymbol{x}))> \delta)]$. For this purpose, The next-best sample is selected at the value of $\boldsymbol{x}$ which is associated with maximum uncertainty in the integrand of \eqref{VHGPEP} (so that the sample is expected to reduce the uncertainty of \eqref{VHGPEP} significantly): 
\begin{align}
    \boldsymbol{x}^* = {\rm{argmax}}_{\boldsymbol{x}} \; {\rm{std}}_{f,g}[\mathbb{P}(S(\boldsymbol{x}, \omega |  f(\boldsymbol{x}), g(\boldsymbol{x})) > \delta)]p_X(\boldsymbol{x}). \label{acq}
\end{align}

We note that \eqref{acq} is closely related to the so-called U criterion \citep{echard2011ak} widely used in computing the exceeding probability associated with a deterministic ItR. In general, the U criterion seeks the most `dangerous' point (i.e., point with maximum local variance), which in our problem corresponds to $\boldsymbol{x}^* = {\rm{argmax}}_{\boldsymbol{x}} \; {\rm{std}}_{f,g} [\mathbb{P}(S(\boldsymbol{x}, \omega |  f(\boldsymbol{x}), g(\boldsymbol{x})) > \delta)]$. Therefore, our acquisition function in \eqref{acq} can be considered as a weighted U criterion which incorporates the influence of the input distribution $p_X(\boldsymbol{x})$ in computing the variance of \eqref{VHGPEP}. Furthermore, the criterion in \eqref{acq} corresponds to the upper bound of ${\rm{var}}_{f,g} [\mathbb{P}(S(X, \omega |  f(\boldsymbol{x}), g(\boldsymbol{x}))> \delta)]$, as we can show
\begin{align}
    & {\rm{var}}_{f,g} [\mathbb{P}(S(X, \omega |  f(\boldsymbol{x}), g(\boldsymbol{x})) > \delta)] 
\nonumber \\
    \leq \; & \frac{1}{2}  \int {\rm{std}}_{f,g}  \Big[ \mathbb{P} (S(\boldsymbol{x}, \omega |  f(\boldsymbol{x}), g(\boldsymbol{x})) > \delta ) \Big] p_X(\boldsymbol{x}) d\boldsymbol{x}. \label{upb}
\end{align}
The derivation for this upper bound is shown in Appendix B.

In practice, we approximate the operator ${\rm{std}}_{f,g} \equiv ({\rm{var}}_{f,g} )^{0.5}$ in \eqref{acq} by the two-dimensional spherical cubature integration \citep{wan2000unscented} with $4$ quadrature points (although extension to more quadrature points is straightforward):
\begin{align}
     & \mathbb{E}_{f,g}[\mathbb{P} (S(\boldsymbol{x}, \omega |  f(\boldsymbol{x}), g(\boldsymbol{x})) > \delta)] 
\nonumber \\ 
    = & \int \mathbb{P}  (S(\boldsymbol{x}, \omega |  f(\boldsymbol{x}), g(\boldsymbol{x})) > \delta) p_{f,g}(f(\boldsymbol{x}), g(\boldsymbol{x})) d f(\boldsymbol{x}) d g(\boldsymbol{x})
\nonumber \\
     \approx & \;\frac{1}{4} \sum_{i=1}^{4} \mathbb{P}(S(\boldsymbol{x}, \omega | \boldsymbol{u}^{(i)} ) > \delta) = m,  
\label{mean} \\
    & {\rm{var}}_{f,g}[\mathbb{P}  (S(\boldsymbol{x}, \omega |  f(\boldsymbol{x}), g(\boldsymbol{x})) > \delta)] 
\nonumber \\
    = & \int (\mathbb{P} (S(\boldsymbol{x}, \omega |  f(\boldsymbol{x}), g(\boldsymbol{x})) > \delta)-m)^2 p_{f,g}(f(\boldsymbol{x}), g(\boldsymbol{x})) d f(\boldsymbol{x}) d g(\boldsymbol{x}) 
\nonumber \\
    \approx & \; \frac{1}{4} \sum_{i=1}^{4} (\mathbb{P}(S(\boldsymbol{x}, \omega | \boldsymbol{u}^{(i)} ) > \delta) - m)^2,  
\label{var} \\
    \boldsymbol{u}^{(1)} = & \; \{f(\boldsymbol{x}) = \mu_f(\boldsymbol{x}) + \sqrt{2 \; {\rm{cov}}_f(\boldsymbol{x})}, \;g(\boldsymbol{x}) = \mu_g(\boldsymbol{x}) \} 
\nonumber\\
    \boldsymbol{u}^{(2)} = & \; \{f(\boldsymbol{x}) = \mu_f(\boldsymbol{x}) - \sqrt{2 \; {\rm{cov}}_f(\boldsymbol{x})}, \;g(\boldsymbol{x}) = \mu_g(\boldsymbol{x}) \} 
\nonumber\\
    \boldsymbol{u}^{(3)} = & \; \{f(\boldsymbol{x}) = \mu_f(\boldsymbol{x}), \; g(\boldsymbol{x}) = \mu_g(\boldsymbol{x}) + \sqrt{2 \; {\rm{cov}}_g(\boldsymbol{x})}\} 
\nonumber\\
    \boldsymbol{u}^{(4)} = & \; \{f(\boldsymbol{x}) = \mu_f(\boldsymbol{x}), \; g(\boldsymbol{x}) = \mu_g(\boldsymbol{x}) - \sqrt{2 \; {\rm{cov}}_g(\boldsymbol{x})}\} 
\label{u},
\end{align}
where the quadrature points \eqref{u} and the corresponding weights $1/4$ in \eqref{mean} and \eqref{var} are selected for third-order accuracy of the scheme (see Appendix C). With \eqref{mean}-\eqref{u} to compute the operator ${\rm{std}}_{f,g} $, \eqref{acq} can be directly solved using standard optimization methods, say a multiple-starting L-BFGS-B quasi-Newton method \citep{nocedal1980updating}.

We remark that the construction of acquisition function has been studied extensively in the case of deterministic ItR \citep{echard2011ak,hu2016global,zhu2016reliability,wang2016gaussian,bichon2008efficient}, and that there may still be room for improvement relative to \eqref{acq} in the case of stochastic ItR. These potential improvements of acquisition function may generally take consideration of correlation between different $\boldsymbol{x}$ in addition to the standard deviation in \eqref{acq}. For example, techniques developed for deterministic ItR, such as using a hypothetical point \citep{farazmand2019extreme,pandita2019bayesian} and global sensitivity analysis \cite{hu2016global}, may be transferred here. However, they may lead to significantly increased computational cost when combining with VHGPR (e.g., the re-training of the variational parameters when hypothetical points are used). These potential developments will be left to our future work.

Combining the VHGPR surrogate model (Eq. \eqref{VHGP1}, \eqref{VHGP2}, and \eqref{VHGP3}) and the optimization of acquisition function \eqref{acq}, we are able to sequentially select the next-best samples starting from an initial dataset. 
The final estimation of exceeding probability $P_e$ is computed by VHGPR with $\mu_f$ and $\mu_g$ to represent functions $f$ and $g$:
\begin{equation}
    P_e = \mathbb{P}\big(S\boldsymbol{(}X, \omega | f(\boldsymbol{x}) = \mu_f(\boldsymbol{x}), g(\boldsymbol{x}) = \mu_g(\boldsymbol{x})\boldsymbol{)} > \delta \big).
    \label{final}
\end{equation}

We summarize the algorithm of this sequential sampling process in Algorithm 1. 

%It should be noted that in constructing the acquisition, we do not use all information we obtained from the VHGPR surrogate, especially the correlation between each prediction. It is anticipated that a full information-based acquisition or its simplified version by measuring the benefits of adding a hypothetical point is better. However, this would greatly increase the computational cost. In VHGPR, the parameters to be optimized in training process are associated with each sample, thus the adding of a hypothetical sample to an existing dataset (one iteration of an optimization process in determining the next sample) requires the retraining of the surrogate. A better acquisition with the acceptable cost is an important future work.

% For SGP, the hyper-parameters are global for all training points. Retraining is not needed after adding a hypothetical sample to the training set, while for VHGP, the hyper-parameters are local for each training point. Thus the adding of a new point to the training set requires retraining of the VHGP.

\begin{algorithm}
    \caption{Sequential sampling for systems with stochastic ItR}
  \begin{algorithmic}
    \REQUIRE Number of initial points $n_{init}$, number of iterations $n_{iter}$
    \INPUT Initial dataset $\mathcal{D}=\{\boldsymbol{x}^i,y^i\}_{i=1}^{i=n_{init}}$
    \STATE \textbf{Initialization} $j = 0$
    \WHILE{$j < n_{iter}$}
      \State Train the surrogate model (Eq. \eqref{VHGP1}, \eqref{VHGP2}, and \eqref{VHGP3}) with $\mathcal{D}$
      \STATE Maximize the acquisition function \eqref{acq} to find the next best sample $\boldsymbol{x}^{j+1}$
      \STATE Implement numerical simulation to get $y^{j+1} = S(\boldsymbol{x}^{j+1}, \omega)$
      \STATE Update the dataset  $\mathcal{D} = \mathcal{D} \cup \{\boldsymbol{x}^{j+1},y^{j+1}\}$
      \STATE $j = j + 1$
    \ENDWHILE
\OUTPUT Compute the exceeding probability  \eqref{final} based on the surrogate model
  \end{algorithmic}
\label{al}
\end{algorithm}

\section{Results}
In this section, we validate our approach using two synthetic problems and a realistic engineering application to quantify the extreme ship motion probability in irregular waves. The heteroscedastic randomness in the ItR are assigned artificially in the former cases, while resulted naturally from dimension reduction of the input parameter space in the latter case. For all cases, we present the results from our current method of sequential sampling (or sequential BED) with VHGPR as a surrogate model (Seq-VHGPR), as well as other methods for validation and comparison. These other methods include Monte Carlo sampling using one million samples for accurate estimation of the mean and variance of ItR (Exact-MC, which serves as the exact result to validate Seq-VHGPR);
% One million samples for X with exact variance 
space-filling Latin hypercube (LH) sampling \cite{mckay2000comparison} with VHGPR as a surrogate (LH-VHGPR, which serves as a reference to demonstrate the efficiency of sequential sampling); LH sampling with SGPR as a surrogate (LH-SGPR, to demonstrate the necessity of using VHGPR). We also include the asymptotic value obtained from the LH-SGPR method, i.e., the convergent result with sufficiently large number of samples. This represents the best solution that can be achieved by previous vast methods based on SGPR with constant uncertainties \citep{hu2016global,echard2011ak,bichon2008efficient,sun2017lif}.

\subsection{One-dimensional (1D) synthetic problem}
\begin{figure}
    \centering
    \includegraphics[width = 8cm]{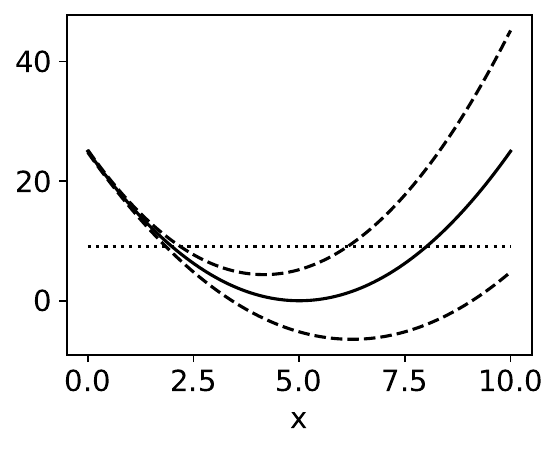}
    \caption{The mean $f(x)$ (\blackline) and uncertainty bounds $f(x) \pm 2\gamma(x)$ (\blackdashedline) of the 1D ItR, as well as the threshold (\blackdottedline) in defining the exceeding probability.}
    \label{fig:f1}
\end{figure}

We start from a 1D synthetic problem, where the true ItR $S(x,\omega)$ \eqref{ItR2} is constructed with (see figure \ref{fig:f1} for an illustration)
\begin{equation}
    f(x) = (x-5)^2,
\end{equation}
and $R(x,\omega)\sim \mathcal{N}(0, \gamma^2(x))$ with
\begin{equation}
    \gamma(x) = 0.1 + 0.1 x ^{2}.
\end{equation}
The input $X$ is assumed to follow a Gaussian distribution with $p_X(x)=\mathcal{N}(5,1)$. Our objective is to estimate an exceeding probability $P_e = \mathbb{P}(S(X,\omega)>9)$. For Seq-VHGPR, we use 40 LH samples as the initial data set, and show the results after 40 initial samples along with other methods.  

%The input $X$ is a Gaussian centered at $5$ with standard deviation 1: $X \sim \mathcal{N}(5,1)$. Our objective is to estimate $\mathbb{P}(S(X,\omega)>9)$. Besides the ItR, we also demonstrate the local contribution of the exceeding probability, $\mathbb{P}(S(x,\omega)>9) p_X(x)$ in 1D ItR (Fig. \ref{fig:f1} (b)), whose integration is the exceeding probability. The most dangerous region which contributes $75\%$ to the exceeding probability is around $x = 7$.

\begin{figure}
    \centering
    \includegraphics[width=10cm]{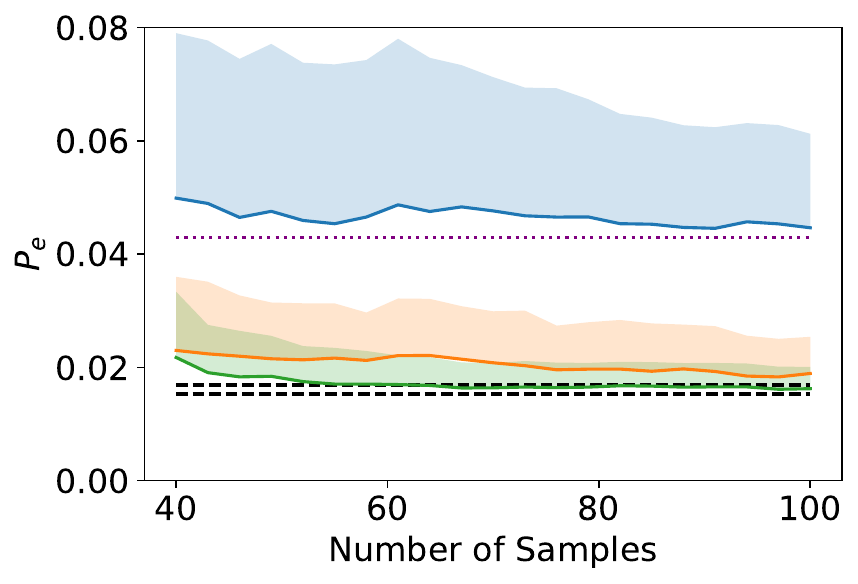}
    \caption{$P_e$ in the 1D synthetic problem, computed by Seq-VHGPR(\greenline), LH-VHGPR(\orangeline), LH-SGPR(\blueline) with its asymptotic value (\purpledottedline), Exact-MC(\blackdashedline) (in terms of the upper and lower 5\% error bounds). The shaded region represents one standard deviation above the mean estimated from 100 applications of the each method.}
    \label{fig:results1}
\end{figure}

\begin{figure}
    \centering
    \includegraphics[width = 12cm]{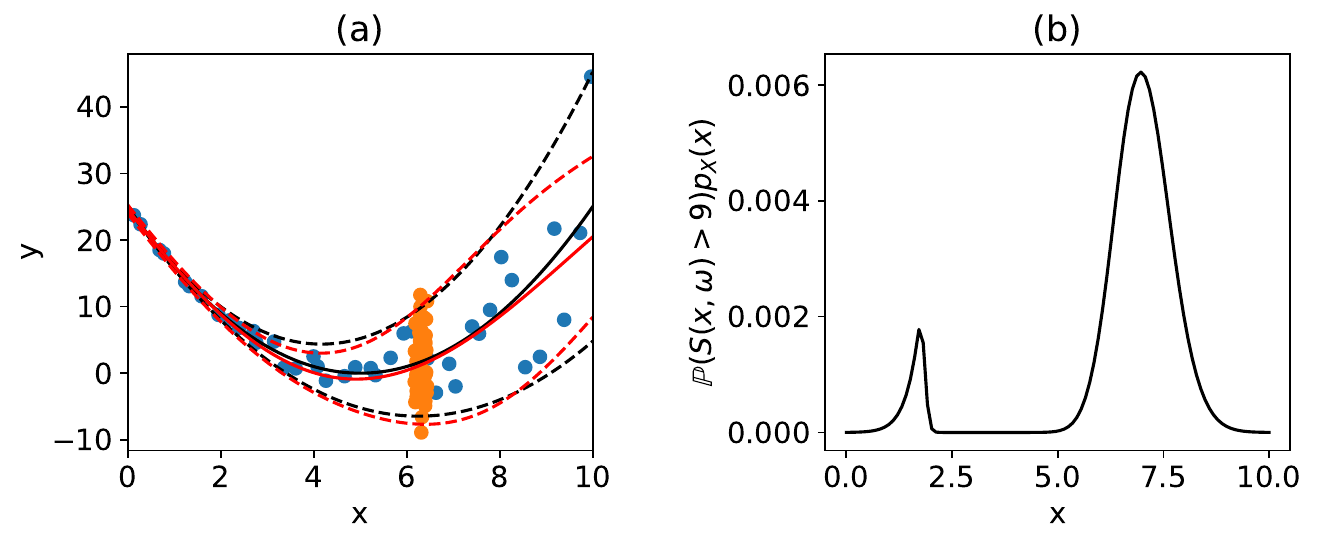}
    \caption{(a) Typical positions of initial 40 samples (\tikzcircle{2pt, NavyBlue}) and 60 sequential samples (\tikzcircle{2pt,Orange}) in Seq-VHGPR, as well as the learned function $f(x)$(\redline) and $f(x) \pm 2\gamma(x)$(\reddashedline) compared to the corresponding exact functions (\blackline, \blackdashedline); (b) the function $\mathbb{P}(S(x,\omega)>\delta) p_X(x)$.}
     \label{fig:learned1} 
\end{figure}

Figure \ref{fig:results1} plots the results $P_e$ computed by Exact-MC, Seq-VHGPR, LH-VHGPR and LH-SGPR (where in Seq-VHGPR and LH-VHGPR, $P_e$ is estimated by \eqref{final}; in LH-SGPR, $P_e$ is estimated by \eqref{final} with constant variance $g(x)=\log(\gamma_0^2)$). Also included in figure \ref{fig:results1} are the standard deviations in Seq-VHGPR, LH-VHGPR and LH-SGPR obtained from 100 applications of the methods. (These uncertainties come from the initial sampling positions and the randomness $\omega$ of ItR in computing $S(x,\omega)$ for each query.) We see that the result from Seq-VHGPR converges rapidly to that from Exact-MC (shown in terms of the 5\%-error region) in the first 20 sequential samples, with an accurate estimation of the exceeding probability. In contrast, the LH-VHGPR result converges much slower, with a non-negligible difference from the Exact-MC result at the end of 100 samples in figure \ref{fig:results1} (in spite of a favorable trend). Furthermore, the LH-SGPR result converges to an asymptotic value which is 3 times of the Exact-MC result, showing the incapability of this class of methods (i.e., most previous methods using SGPR) in estimating the exceeding probability induced by a heteroscedastic ItR. We remark that the failure of the SGPR-based methods lie in the loss of heteroscedasticity information in ItR, irrespective of the sampling approach or acquisitions used. Finally, as shown by the shaded area in figure \ref{fig:results1}, Seq-VHGPR leads to significantly reduced standard deviation compared to other approaches.

We further examine the reason for the fast convergence achieved by Seq-VHGPR (relative to LH-VHGPR). Figure \ref{fig:learned1}(a) plots the positions of 100 samples (i.e., 40 initial and 60 sequential samples) in Seq-VHGPR, as well as the learned functions $f(x)$ and $\gamma(x)$. While the initial 40 samples are randomly chosen (providing the overall trend of $f(x)$ and $\gamma(x)$), the 60 sequential samples are concentrated near $x=6.5$, providing more accurate estimation of $f(x)$ and $\gamma(x)$ in the nearby region. This point corresponds to the maximum in $\mathbb{P}(S(x, \omega)>\delta) p_X(x)$ (the integrand in \eqref{exprob}) as shown in figure \ref{fig:learned1}(b), leading to the largest contribution in computing the exceeding probability \eqref{exprob}.

%The results (Fig. \ref{fig:results1}) show that our sequential algorithm delivers better performance than random sampling with SGPR and VHGPR. On the one hand, the SGPR-based methods demonstrate slower convergence and large discrepancy with the exact. Actually, even with an infinite number of samples, it is still hard for SGPR-based algorithm to approach true statistics due to its over-simplified constant-randomness model (note the difference between Asy-SGPR and Exact-Sto). On the other hand, compared with LH sampling, the sequential samples generally concentrated on dangerous region which contribute mostly to exceeding probability (Fig. \ref{fig:learned1}). This concentration helps VHGPR learn the mean and (more difficult) variance of dangerous region (Fig. \ref{fig:learned1}) with high accracy than SGPR based method, leading to a better estimation of exceeding probability. 

\subsection{Two-dimensional (2D) synthetic problem}

We construct a 2D synthetic problem by setting $f(\boldsymbol{x})$ to be a four-branch function (that has been widely-used in estimating extreme-event probability with a deterministic ItR) \citep{echard2011ak,hu2016global,sun2017lif,wang2016gaussian}:
$$
f(x_1, x_2) = - {\rm{min}}\left\{
\begin{aligned}
    & 8+ 0.1(x_1-x_2)^2 + \frac{(x_1 + x_2)}{\sqrt{2}}                     \\
    & 8+ 0.1(x_1-x_2)^2 - \frac{(x_1 + x_2)}{\sqrt{2}}                       \\
    & (x_1 - x_2) + \frac{6}{\sqrt{2}} + 5             \\
    & (x_2 - x_1) + \frac{6}{\sqrt{2}} + 5                    \\
\end{aligned}
\right..
$$

\begin{figure}
    \centering
    \includegraphics[trim=0 0 10.5cm 0,clip,width=12cm]{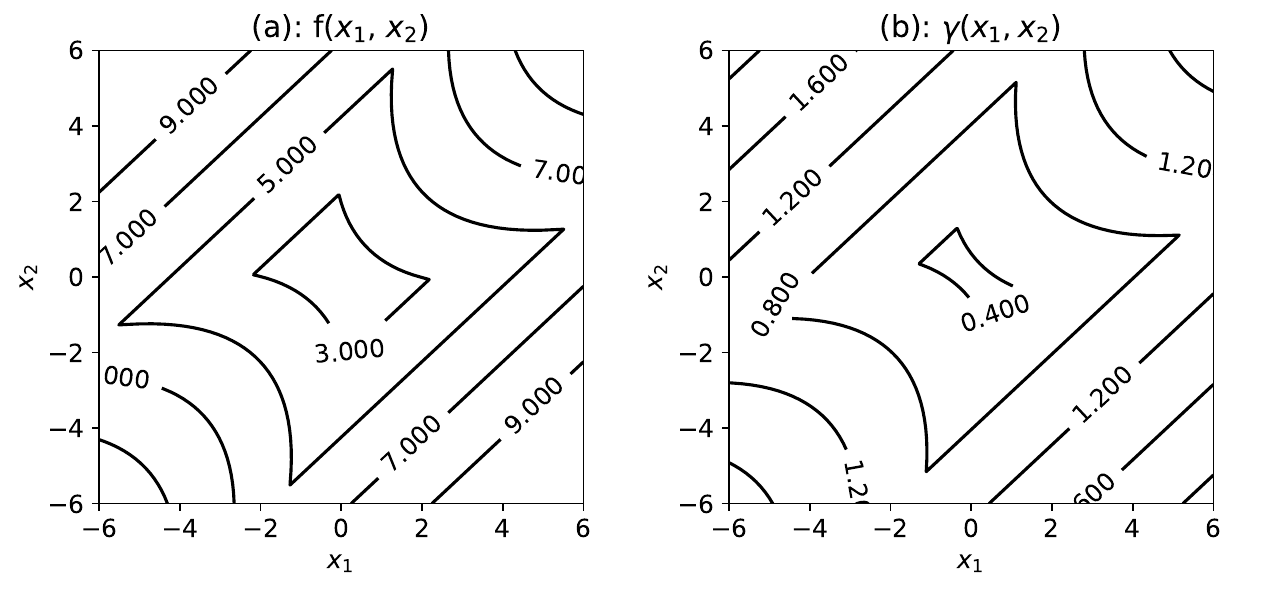}
    \caption{(a) $f(x_1,x_2)$ and (b) $\gamma(x_1,x_2)$ as in $R(x_1,x_2,\omega)$ in the 2D stochastic four-branch ItR.}
    \label{fig:f2}
\end{figure}

To generate an uncertain ItR, we add a Gaussian randomness $R(x_1,x_2,\omega)\sim \mathcal{N}(0, \gamma^2(x_1,x_2))$ to $f(x_1, x_2)$ with standard deviation $\gamma(x_1, x_2) = f(x_1, x_2) / 6 \;$ (see figure \ref{fig:f2} for $f(x_1,x_2)$ and $\gamma(x_1,x_2))$). We assume the input $X$ to follow a Gaussian distribution $p_{X_1,X_2}(x_1,x_2) = \mathcal{N}(\boldsymbol{0}, {\rm{I}})$, with $\rm{I}$ being a 2$\times$2 identity matrix, and our purpose is to estimate an exceeding probability $P_e=\mathbb{P}(S(X_1,X_2, \omega)>5)$. For Seq-VHGPR, 60 LH samples are used as the initial data set.

\begin{figure}
    \centering
    \includegraphics[width=10cm]{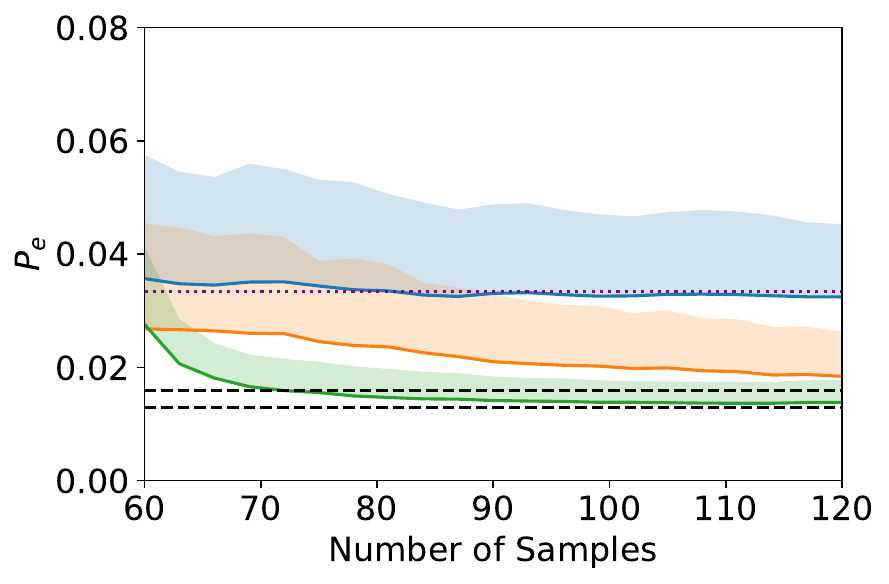}
    \caption{$P_e$ in the 2D synthetic problem, computed by Seq-VHGPR(\greenline), LH-VHGPR(\orangeline), LH-SGPR(\blueline) with its asymptotic value (\purpledottedline), Exact-MC(\blackdashedline) (in terms of the upper and lower 5\% error bounds). The shaded region represents one standard deviation above the mean estimated from 100 applications of the each method.}
    \label{fig:results2}
\end{figure}

\begin{figure}
\centering
\begin{subfigure}{.6\textwidth}
  \centering
  \includegraphics[width=8.1cm]{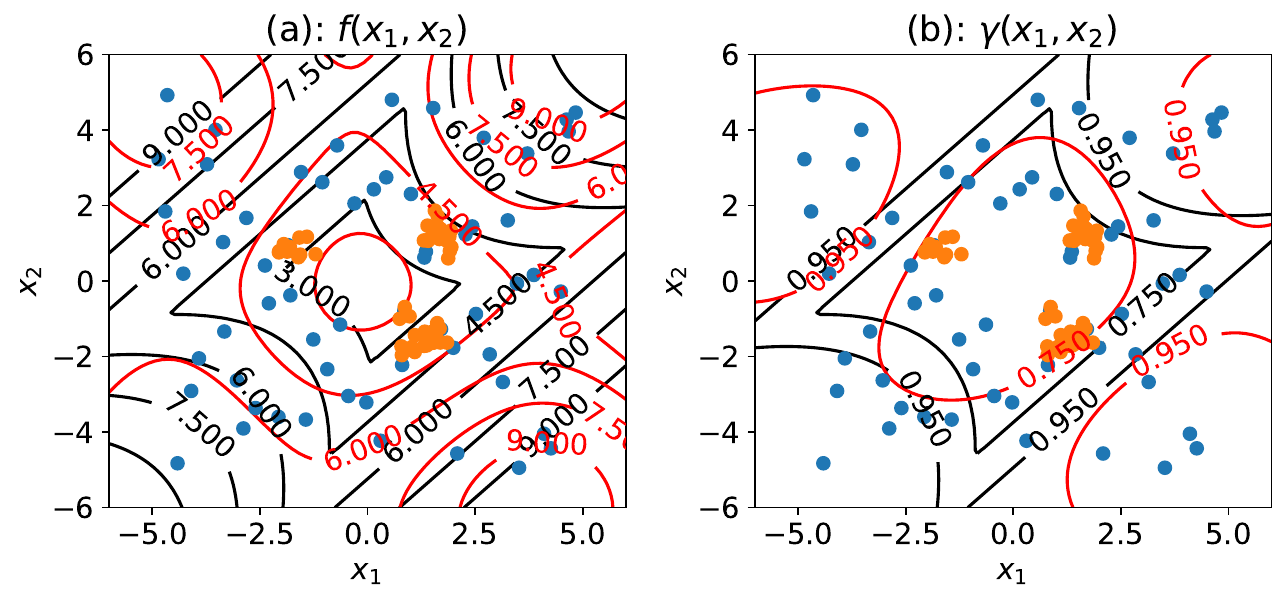}
\end{subfigure}%
\begin{subfigure}{.4\textwidth}
  \centering
  \includegraphics[trim=21.5cm 0 0.2cm 0,clip, width=3.8cm]{figure/2.pdf}
\end{subfigure}
\caption{Typical positions of initial 60 samples (\tikzcircle{2pt, NavyBlue}) and 60 sequential samples (\tikzcircle{2pt,Orange}) in Seq-VHGPR, as well as the learned $f(x_1,x_2)$ (\redline) compared to the exact function (\blackline) in (a); and the learned $\gamma(x_1,x_2)$ (\redline) compared to the exact function (\blackline) in (b);
(c) the (log) function $\mathbb{P}(S(x_1,x_2,\omega)>\delta) p_{X_1,X_2}(x_1,x_2)$.}
\label{fig:learned2}
\end{figure}

The results of the 2D problem, as shown in figure \ref{fig:results2}, further demonstrates the effectiveness of Seq-VHGPR, which approaches the Exact-MC solution with 5\% error within the first 20 sequential samples and leads to the smallest uncertainty among all methods. The convergence rate of Seq-VHGPR is much faster than that of LH-VHGPR, where the latter fails to converge at the end of 120 samples. Compared with 1D results, the superiority of Seq-VHGPR over LH-VHGPR is more evident due to the increased sparsity of samples in the 2D case. The LH-SGPR result, on the other hand, converges to an asymptotic value which is 2.5 times of the Exact-MC result, a significant error due to the neglect of heteroscedastic randomness in ItR.  

The typical positions of (60 initial and 60 sequential) samples in Seq-VHGPR are shown in figure \ref{fig:learned2}(a)(b), as well as the the learned functions $f(x_1,x_2)$ and $\gamma(x_1,x_2)$. Similar to the 1D case, the sequential samples are expected to concentrate in regions where $\mathbb{P}(S(x_1,x_2, \omega)>\delta) p_{X_1,X_2}(x_1,x_2)$ is maximized, i.e., the four regions enclosed by $-4$ contour lines in figure \ref{fig:learned2}(c). As shown in figure \ref{fig:learned2}(a)(b), most sequential samples lie in three out of the four regions (although the situation depends on the initial samples and for some cases all four regions can be filled). The difficulty of the sequential samples transiting to all four regions within 60 samples can be anticipated, which is consistent with the observation in the case of deterministic ItR if the U criterion is used as the acquisition function \citep{echard2011ak}. While the design of better acquisition function is possible referring to the counterpart in the deterministic case \citep{hu2016global}, the current results already show the adequacy of Seq-VHGPR in estimating the exceeding probability (even if not all four regions are filled and the estimation of $\gamma(x_1,x_2)$ is relatively less accurate than that of $f(x_1,x_2)$). 

% 1. The vhgpr performs better with same number of samples
% 2. The vhgpr needs a large number of samples to obtain a roughly good approximation

%As before, we plot the positions of samples (i.e., 60 initial and 60 sequential samples) in Seq-VHGPR, as well as the learned functions $f(x)$ (figure \ref{fig:learned2}(a)) and $\gamma(x)$ (figure \ref{fig:learned2}(b)). In 2D problem, generally 3 in 4 important regions (four ellipses enclosed by -4 log contour lines of $\mathbb{P}(S(\boldsymbol{x}, \omega)>\delta) p_X(\boldsymbol{x})$ in figure \ref{fig:learned2}(c)) can be discovered by sequential samples. The finding (and the subsequent learning) of all dangerous regions becomes challenging in 2D four-branches problem, which is anticipated as even the learning of the limit state (one contour line: $f(x_1, x_2 ) = 5$) requires $O(40)$ samples with the state of art algorithm \citep{hu2016global}. However, our VHGP can still roughly approximate the mean and variance to obtain an acceptable estimation. We also note that the algorithm based on sequential VHGP shows quick convergence at the beginning but becomes sluggish when results enter into the 5 percent error bounds in both 1D and 2D cases. This is due to the fact that an exact inference of the randomness requires a far larger number of samples than the current, e.g. $O(10000)$ in \citep{liu2020large} for a 2D problem with comparable topology complexity.

\subsection{Probability of extreme ship motion in irregular waves}

\begin{figure}
    \centering
    \includegraphics[width =12cm]{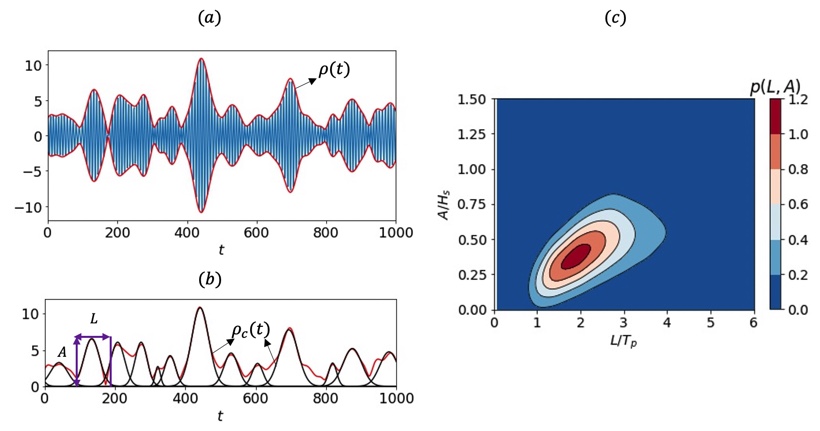}
    \caption{(a) surface elevation $\eta(t)$ (\blueline) and the corresponding envelope process $\rho(t)$ (\redline) in a narrow-band wave field. (b) $\rho(t)$ (\redline) fitted by an ensemble of Gaussian wave groups $\rho_c(t)$(\blackline) with parameters $L$ and $A$.
    (c) $p_{L,A}(L,A)$ obtained from wave fields.}
    \label{fig:wavefield}
\end{figure}

We further consider an engineering application of our method to estimate the probability of extreme ship roll motions in uni-directional irregular waves. In marine engineering, the ship motion problem can often be treated as a dynamical system where the input is a time series of wave (or surface) elevation $\eta(t)$, and the output is, say, the ship roll motion $\xi(t)$. The ItR connecting $\eta(t)$ and $\xi(t)$ can be computed by Computational Fluid Dynamics (CFD) simulations. However, the resolution of exact exceeding probability requires running expensive CFD simulations with a very long-time input $\eta(t)$ (due to the rareness of the extreme roll motion), leading to prohibitively high computational cost. Therefore, for the purpose of validating our approach, we use an inexpensive phenomenological nonlinear roll equation \citep{umeda2004nonlinear} to construct the ItR (with the uncertainty associated with $\omega$ introduced later)
\begin{equation}
     \ddot{\xi}+\alpha_1 \dot{\xi}+\alpha_2 \dot{\xi}^3 +(\beta_1+\epsilon_1 \cos (\phi) \eta(t) )\xi+\beta_2 \xi^3 = \epsilon_2 \sin (\phi) \eta(t), 
    \label{roll}
\end{equation}
with empirical coefficients \citep{nayfeh1980nonlinear} $\alpha_1=0.19$, $\alpha_2=0.06$, $\beta_1=0.04$, $\beta_2=-0.1$, $\epsilon_1=0.020$, $\epsilon_2=0.004$, and $\phi = \pi/6$.

The wave elevation $\eta(t)$ is usually specified from a wave spectral process, which resides in a high-dimensional input space. A typical procedure to reduce the dimension is to describe $\eta(t)$ by an ensemble of wave groups embedded in its envelope process \citep{longuet1983joint,farazmand2019extreme} (see figure \ref{fig:wavefield}(a) for an illustration). Specifically, we compute the envelope process $\rho(t)$ from $\eta(t)$ through the Hilbert transform \citep{shum1984estimates}, and then construct two-parameter Gaussian-like wave groups $\rho_c(x)$ which best fits $\rho(t)$: 
\begin{equation}
    \rho_c(t)\sim A \exp \frac{-(t-t_c)^2}{2L^2},
    \label{GauWG}
\end{equation}
where $t_c$ is the temporal location of the group, and the two parameters $A$ (group amplitude) and $L$ (group length) describe the geometry of the group (figure \ref{fig:wavefield}(b)). This dimension-reduction procedure allows $\eta(t)$ to be described by an ensemble of $(L,A)$ wave groups, i.e., a two-dimensional input parameter space (see figure \ref{fig:wavefield}(c)). We can then construct an ItR with the input as $(L,A)$ to \eqref{roll} and the output as the maximum roll through this wave group, and consider the group-based probability.

%Sampling in the low-dimensional space gives us tremendous computational saving, which has been used in \cite{mohamad2018sequential} and \cite{gong2020}. In these works, the responses of a ship in a group with $A$ and $L$ are determined by simulations using standard Gaussian-like envelopes, pre-defined phase and initial response. In this way, the ItR is deterministic. 

\begin{figure}
    \centering
    \includegraphics[width = 12cm]{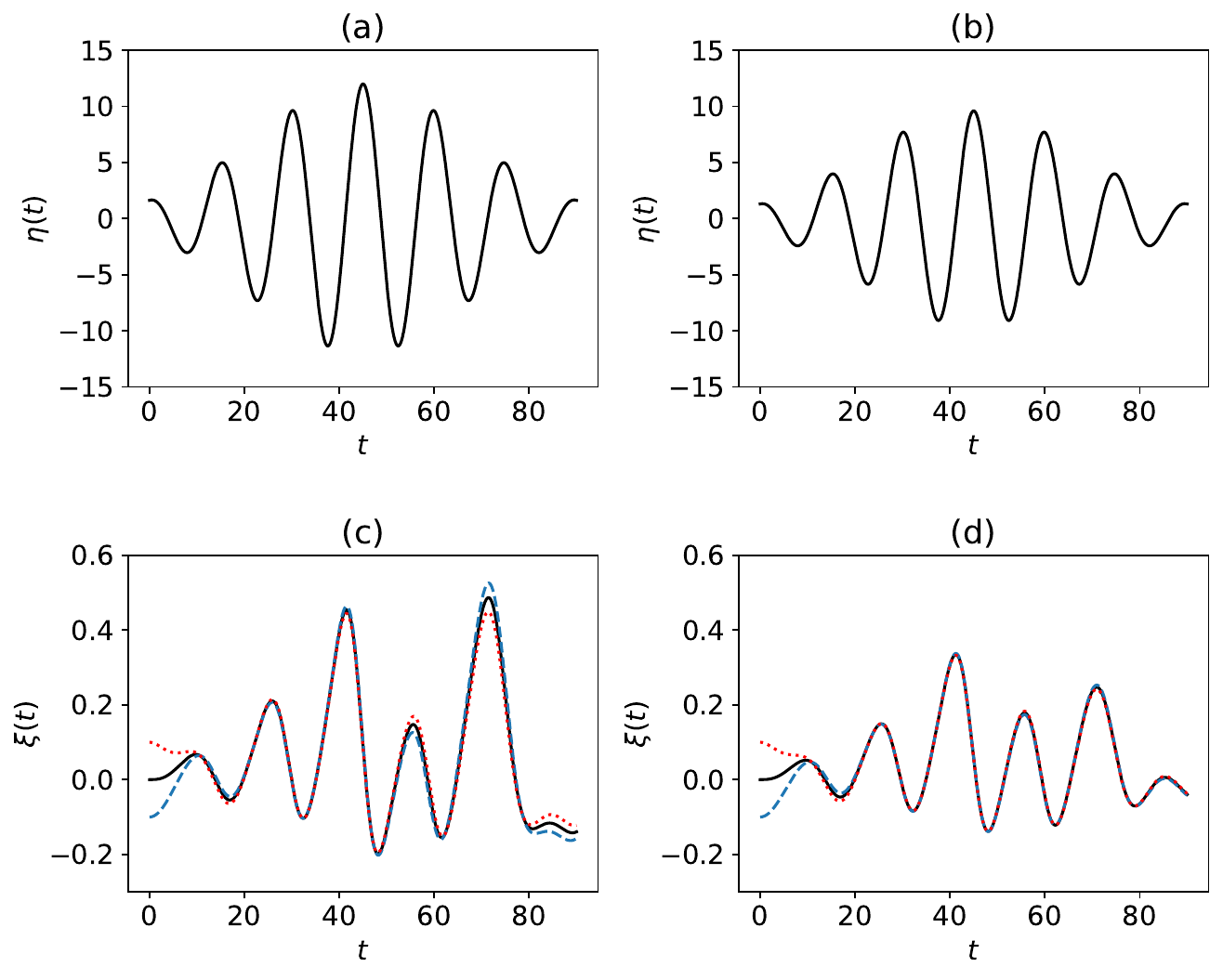}
    \caption{The roll responses (c) and (d) with different initial conditions $\{\dot{\xi}(0) = 0, \xi(0) =0\}$ (\blackline); $\{\dot{\xi}(0) = 0, \xi(0) =-0.05\}$ (\bluedashedline); $\{\dot{\xi}(0) = 0, \xi(0) =0.05\}$ (\reddottedline), computed from \eqref{roll} with input from a wave group with respectively (a) larger and (b) smaller amplitudes.}
    \label{fig:initcon}
\end{figure}

However, the dimension reduction results in the loss of information relative to the original field $\eta(t)$, i.e., it introduces uncertainties in the ItR, including the uncertain initial conditions of $(\xi(0), \dot{\xi}(0))$ and detailed phase and frequency conditions in the wave group.  As shown in \citep{anastopoulos2019evaluation,soliman1991transient}, \eqref{roll} (and the ship roll in general) may be sensitive to the lost information such as initial conditions, and the resulted uncertainty is non-uniform for different $A$ and $L$.  (see figure \ref{fig:initcon} as an example that the uncertainty is larger for the first wave group but smaller for the second one). This creates heteroscedastic uncertainty in the ItR (associated with $\omega$) that needs to be dealt with by our current approach Seq-VHGPR. 

%Besides, the phase which may change within a group and discrepancy between true group and perfect Gaussian shape should also be considered.  

In the following, we show the results with input $\eta(t)$ extracted from a narrow-band Gaussian wave spectrum:
\begin{equation}
F(k) =  \frac{H_s^2}{16}\frac{1}{\sqrt{2\pi}\mathcal{K}}\exp \frac{-(k-k_0)^2}{2\mathcal{K}^2},  
\label{Fk}
\end{equation}
with significant wave height $H_s=12m$, peak (carrier) wavenumber $k_0=0.018 m^{-1}$ (corresponding to peak period $T_p=15s$), and $\mathcal{K}=0.05k_0$. The exact exceeding probability is computed by simulating $1500$ hours (360000 $T_p$) of ship responses. To compute the ItR incorporating the heteroscedastic randomness in $\omega$, after a sample $(L,A)$ is chosen, we randomly select a wave group of this $(L,A)$ in $\rho(x)$, and simulate \eqref{roll} starting from (on average) 3 groups ahead of the $(L,A)$ group with a $(0,0)$ initial condition. Since the impact of initial conditions typically decay in $O(1)$ wave group, we are able to naturally capture the true initial condition as the ship encounters the $(L,A)$ group, as well as the phase and frequency condition in the particular group.  

\begin{figure}
    \centering
    \includegraphics[width=10cm]{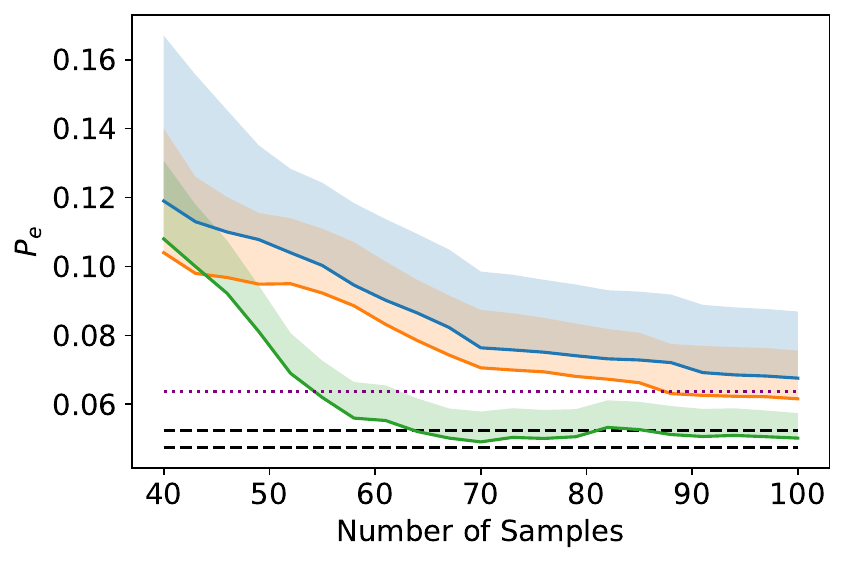}
    \caption{$P_e$ in the ship roll problem, computed by Seq-VHGPR(\greenline), LH-VHGPR(\orangeline), LH-SGPR(\blueline) with its asymptotic value (\purpledottedline), exact solution(\blackdashedline) (in terms of the upper and lower 5\% error bounds). The shaded region represents one standard deviation above the mean estimated from 50 applications of the each method.}
    \label{fig:results3}
\end{figure}

Figure \ref{fig:results3} plots  $P_e= \mathbb{P}({\rm{max}}( \xi_{L,A}(t)) > 0.3)$ (the probability that maximum roll in a group exceeds 17 degrees) obtained from Seq-VHGPR, LH-VHGPR, LH-SGPR and the exact solution. We see that the Seq-VHGPR result converges to the exact solution within the first 30 sequential samples, much faster than the convergence of the LH-VHGPR result. The LH-SGPR result converges to an asymptotic value which is appreciably larger than the exact solution. We remark that this value represents the best result that can be achieved by all previous methods on this problem \citep{mohamad2018sequential,gong2021full} based on SGPR. These results, again, demonstrate the effectiveness of Seq-VHGPR in computing the exceeding probability relative to all other approaches. 

\begin{figure}
    \centering
    \includegraphics[width = 12cm]{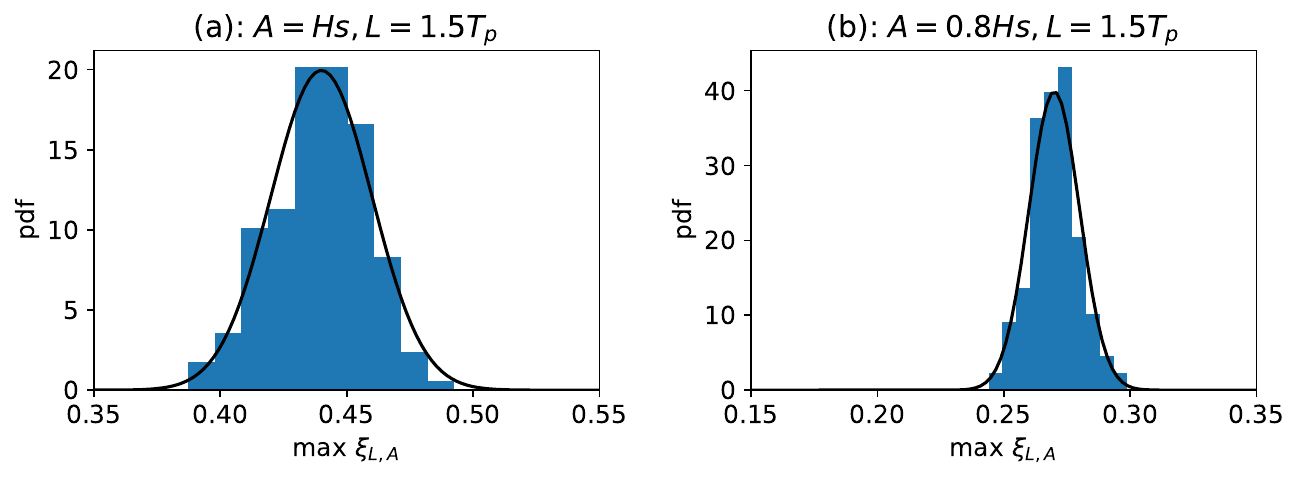}
    \caption{The density histograms of $S(A, L, \omega)$ for (a) $A=H_s,L=1.5T_p$ and (b) $A=0.8H_s,L=1.5T_p$, generated from all such groups in the time series of five million $T_p$. The Gaussian fits for the histograms are shown (\blackline).}
    \label{fig:Gaussian_randomness}
\end{figure}

Finally, the application of VHGPR in our problem assumes that the distribution of $S(\boldsymbol{x},\omega)$ is approximately Gaussian (associated with $\omega$ for each $\boldsymbol{x}\equiv (A,L)$). While this cannot be checked in the Seq-VHGPR sampling, we provide a posterior calculation to show that this is indeed true. Figure \ref{fig:Gaussian_randomness} plots the distribution of $S(\boldsymbol{x},\omega)$ for two selected values of $(A=H_s,L=1.5T_p)$ and $(A=0.8H_s,L=1.5T_p)$, generated from all such groups in the time series of five million $T_p$. It is evident that the distributions are approximated by Gaussian distributions.

\section{Conclusions}
In this paper, we present a new method (Seq-VHGPR) to efficiently estimate the extreme-event probability (in terms of the exceeding probability) induced by an ItR with heteroscedastic uncertainty. The method is established in the framework of sequential BED, and leverages the VHGPR as a surrogate model to estimate the uncertain ItR. A new acquisition function corresponding to the VHGPR estimation is developed to select the next-best sequential sample which leads to fast convergence of the exceeding probability. We validate our new method in two synthetic problems and one engineering application to estimate the extreme ship motion probability in irregular waves. In all cases, we find fast convergence of Seq-VHGPR to the exact solution, demonstrating its superiority to all existing methods if an ItR with heteroscedastic uncertainty is associated with the problem. This is indeed due to the effectiveness of VHGPR in estimating the ItR, although there is still room for improvement of the acquisition function to accelerate the convergence as done in the case of deterministic ItR \citep{hu2016global}. Finally, we remark that the present method also provides an effective way for high-dimensional BED, where the most influential dimensions can be selected as (low-dimensional) input $X$, with other secondary ones packaged into $\Omega$ in the ItR (as in the ship motion problem).

\section*{ACKNOWLEDGEMENT}

This research is supported by the Office of Naval Research
grant N00014-20-1-2096. We thank the program manager Dr. Woei-Min Lin for several helpful discussions
on the research. 
%This research is supported by Office of Naval Research grant N00014-20-1-2096. We thank the program manager Dr. Woei-Min Lin for several helpful discussions on the research. 
This work used the Extreme Science and Engineering Discovery Environment (XSEDE) [Towns 2014] through allocation TG-BCS190007.

\bibliographystyle{elsarticle-num} 
\bibliography{reference.bib}

\appendix

\section{Algorithms for Gaussian Process Regression}
We consider the task of inferring the input to response (ItR) function from a dataset $\mathcal{D}=\{\boldsymbol{x}^i,y^i\}_{i=1}^{i=n}$ consisting of $n$ inputs $\mathbb{X} = \{\boldsymbol{x}^{i}\in \mathbb{R}^d \}_{i=1}^{i=n}$ and the corresponding outputs $\boldsymbol{y} = \{y^{i}\in \mathbb{R}\}_{i=1}^{i=n}$.
\subsection{Standard Gaussian Process Regression (SGPR)}
 SGPR assumes the function to be a sum of a mean $f(\boldsymbol{x})$ and a Gaussian randomness with constant variance $\gamma_0^2$ (at all $\boldsymbol{x}$):
\begin{equation}
    y = f(\boldsymbol{x}) + R  \quad R \sim \mathcal{N}(0, \gamma_0^2),
\end{equation}
A prior, representing our beliefs over all possible functions we expect to observe, is placed on $f$ as a Gaussian process $f(\boldsymbol{x}) \sim \mathcal{GP}(0,k_f(\boldsymbol{x},\boldsymbol{x}'))$ with zero mean and covariance function $k_f$ (usually defined by a radial-basis-function kernel): 
\begin{equation}
k_f(\boldsymbol{x},\boldsymbol{x}') = \tau^2 {\rm{exp}}(-\frac{1}{2} \sum_{j=1}^{j=d}\frac{(x_j-x_j')^2}{l_j^2} ), \label{RBF}
\end{equation}
where the amplitude $\tau^2$ and length scales $l_j$, together with $\gamma_0$, are hyperparameters $\boldsymbol{\theta}=\{\tau, l_j, \gamma_0\}$ in SGPR.

Following the Bayesian formula, the posterior prediction for $f$ given the dataset $\mathcal{D}$ can be derived to be another Gaussian: 
\begin{equation}
    p(f(\boldsymbol{x})|\mathcal{D}) = \frac{p(f(\boldsymbol{x}),\boldsymbol{y})}{p(\boldsymbol{y})} = \mathcal{N}(\mu_f(\boldsymbol{x}), {\rm{cov}}_f(\boldsymbol{x})),
\end{equation}
with analytically tractable mean $\mu_f(\boldsymbol{x})$ and covariance $ {\rm{cov}}_f(\boldsymbol{x},\boldsymbol{x}')$:
 \begin{align}
    \mu_f(\boldsymbol{x}) & = k_f(\boldsymbol{x}, \mathbb{X})^T ({\rm{K}}_f(\mathbb{X},\mathbb{X})+\gamma_0^2 {\rm{I}})^{-1} \boldsymbol{y}, \label{SGPPredf}\\
    {\rm{cov}}_f(\boldsymbol{x}, \boldsymbol{x}') & = k_f(\boldsymbol{x},\boldsymbol{x}') - k_f(\boldsymbol{x},\mathbb{X})^T ({\rm{K}}_f(\mathbb{X},\mathbb{X})+\gamma_0^2 {\rm{I}})^{-1} k_f(\boldsymbol{x}',\mathbb{X}),\label{SGPPredC}
\end{align}
where matrix element ${\rm{K}}_f(\mathbb{X},\mathbb{X})_{ij}=k_f(\boldsymbol{x}^i,\boldsymbol{x}^j)$. The hyperparameters $\boldsymbol{\theta}$ are determined which maximizes the likelihood function $p(\mathcal{D}|\boldsymbol{\theta})\equiv p(\boldsymbol{y}|\boldsymbol{\theta})=\mathcal{N}(0, {\rm{K}}_f(\mathbb{X},\mathbb{X}) + \gamma_0^2 {\rm{I}})$.

\subsection{Variational Heteroscedastic Gaussian Process Regression (VHGPR)}
In this section, we briefly outline the algorithm of VHGPR. The purpose is to provide the reader enough information to understand the logic behind VHGPR. For the conciseness of the presentation, some details in the algorithm have to be omitted. We recommend the interested readers to read \citep{lazaro2011variational} and Sec. 10 in  \citep{bishop2006pattern} for details.

In VHGPR, the function of ItR is considered as the sum of a mean and a Gaussian term (independent at all $\boldsymbol{x}$) with heteroscedastic uncertainty:
\begin{equation}
    y = f(\boldsymbol{x}) + R  \quad R \sim \mathcal{N}(0, e^{g(\boldsymbol{x})}).
\end{equation}
Two Gaussian priors are placed on the mean and the (log) variance function:
\begin{align}
    f(\boldsymbol{x}) & \sim \mathcal{GP} (0, k_f(\boldsymbol{x},\boldsymbol{x}')), \label{kf}\\ 
    g(\boldsymbol{x}) & \sim \mathcal{GP} (\mu_0, k_g(\boldsymbol{x},\boldsymbol{x}')) \label{kg},
\end{align}
where $k_f$ and $k_g$ are respectively the radial-basis-function kernels for $f$ and $g$, defined similarly as \eqref{RBF}. $\mu_0$ is the prior mean for $g$. The hyperparameters in VHGPR can then be defined as $\boldsymbol{\theta}=(\mu_0, \boldsymbol{\theta}_{f}, \boldsymbol{\theta}_{g})$ where $\boldsymbol{\theta}_{f,g}$ includes the amplitudes and length scales involved in $k_f$ or $k_g$. Moreover, we assume $f$ is independent with $g$.

The increased expressive power with the heteroscedastic variance is at the cost of analytically intractable likelihood (for determination of hyperparameters) and posterior (prediction). 
% 1. likelihood 2. posterior latent distribution 3. posterior prediction distribution
Let $\boldsymbol{f} = f(\mathbb{X})$ and $\boldsymbol{g} = g(\mathbb{X})$ denote the realizations of mean and variance at training inputs $\mathbb{X}$ following the distributions in \eqref{kf} and \eqref{kg} . The likelihood and prediction are formulated as:
\begin{equation}
    p(\mathcal{D}|\boldsymbol{\theta})\equiv p(\boldsymbol{y}|\boldsymbol{\theta}) = \iint p(\boldsymbol{y}|\boldsymbol{f},\boldsymbol{g}) p(\boldsymbol{g}|\boldsymbol{\theta}) p(\boldsymbol{f}|\boldsymbol{\theta}) {\rm{d}} \boldsymbol{g} {\rm{d}} \boldsymbol{f}, \label{llh}
\end{equation}
\begin{equation}
    p(f(\boldsymbol{x}),g(\boldsymbol{x})|\mathcal{D}) = \iint p(f(\boldsymbol{x}),g(\boldsymbol{x})|\boldsymbol{f},\boldsymbol{g}) p(\boldsymbol{f}, \boldsymbol{g}|\boldsymbol{y}) d \boldsymbol{f} d \boldsymbol{g}. \label{pred}
\end{equation}

Since analytical integration cannot be achieved for \eqref{llh} and \eqref{pred}, numerical evaluations of the integrals are needed for their computations. However, the dimension of integration (w.r.t $\boldsymbol{f}$ and $\boldsymbol{g}$) is the same as number of data points $n$, which can be prohibitively high for a direct integration, say, using quadrature methods. While the Monte-Carlo method (e.g., MCMC) offers some advantages in computational cost, its application is still too expensive for most practical problems. For these problems, the VHGPR leveraging variational inference is the only method which provides practical solutions with low computational cost and sufficient accuracy. 
% high dimensional MC. The number of samples are not influenced by dimensions, but how about the generation of the samples? 

The key distribution in computing \eqref{pred} and \eqref{llh} is $p(\boldsymbol{f}, \boldsymbol{g}|\boldsymbol{y})$ (directly involved in \eqref{pred} and related to \eqref{llh} due to \eqref{logllh} that will be discussed), which is however expensive to compute directly. The key idea in VHGPR is to approximate $p(\boldsymbol{f}, \boldsymbol{g}|\boldsymbol{y})$ by $q(\boldsymbol{f}, \boldsymbol{g})$, where the latter is assumed to have a Gaussian distribution with parameters (multi-dimensional means and covariance) denoted here by $\boldsymbol{\theta}_q$. Through minimizing the KL divergence \citep{kullback1951information} between $p(\boldsymbol{f}, \boldsymbol{g}|\boldsymbol{y})$ and $q(\boldsymbol{f}, \boldsymbol{g})$, the parameters $\boldsymbol{\theta}_q$ can be determined and both the posterior and likelihood can be evaluated accordingly as discussed below.

For the posterior, \eqref{pred} becomes a linear Gaussian model (an integration of the  exponential of quadratic function of $\boldsymbol{f}$ and $\boldsymbol{g}$ with Gaussian weights) which has an analytical formulation. For the likelihood, we avoid directly using \eqref{llh} but decompose $\log p(\boldsymbol{y}|\boldsymbol{\theta})$ as a summation of the evidence lower bound (ELBO) $\mathcal{L}(q(\boldsymbol{f}, \boldsymbol{g}))$ and the K-L divergence between $q(\boldsymbol{f}, \boldsymbol{g})$ and $p(\boldsymbol{f}, \boldsymbol{g}| \boldsymbol{y})$ \footnote{This decomposition can be derived from manipulation of  \eqref{KL} to be
\begin{equation}
    {\rm{KL}}(q(\boldsymbol{f}, \boldsymbol{g})|p(\boldsymbol{f}, \boldsymbol{g}| \boldsymbol{y})) = - \iint q(\boldsymbol{f}, \boldsymbol{g}) \log \frac{p(\boldsymbol{y}, \boldsymbol{f},\boldsymbol{g})}{q(\boldsymbol{f}, \boldsymbol{g})} {\rm{d}} \boldsymbol{f} {\rm{d}} \boldsymbol{g} + \log p(\boldsymbol{y}) \nonumber.
\end{equation}
}:
\begin{equation}
    \log p(\boldsymbol{y}|\boldsymbol{\theta}) = \mathcal{L}(q(\boldsymbol{f}, \boldsymbol{g})) + {\rm{KL}}(q(\boldsymbol{f}, \boldsymbol{g})|p(\boldsymbol{f}, \boldsymbol{g}| \boldsymbol{y})),
    \label{logllh}
\end{equation}
where
\begin{align}
    \mathcal{L}(q(\boldsymbol{f}, \boldsymbol{g})) & = \iint q(\boldsymbol{f}, \boldsymbol{g}) \log \frac{p(\boldsymbol{y}, \boldsymbol{f},\boldsymbol{g})}{q(\boldsymbol{f}, \boldsymbol{g})} {\rm{d}} \boldsymbol{f} {\rm{d}} \boldsymbol{g}, \label{ELBO}\\
    {\rm{KL}}(q(\boldsymbol{f}, \boldsymbol{g})|p(\boldsymbol{f}, \boldsymbol{g}| \boldsymbol{y})) & = \iint q(\boldsymbol{f}, \boldsymbol{g}) \log \frac{q(\boldsymbol{f}, \boldsymbol{g})}{p(\boldsymbol{f}, \boldsymbol{g}| \boldsymbol{y})}
    {\rm{d}} \boldsymbol{f} {\rm{d}} \boldsymbol{g} \label{KL}.
\end{align}
We then formulate an optimization problem of $\mathcal{L}$ (where $\mathcal{L}$, as a weighted integration of the exponential function of $\boldsymbol{f}$ and $\boldsymbol{g}$, has an analytical expression for Gaussian weights $q(\boldsymbol{f},\boldsymbol{g})$ \citep{bishop2006pattern}) to determine both the parameters in $q(\boldsymbol{f}, \boldsymbol{g})$ ($\boldsymbol{\theta}_q$) and the hyperparameters ($\boldsymbol{\theta}$):
\begin{equation}
    \boldsymbol{\theta}^*_q, \boldsymbol{\theta}^* = {\rm{argmax}}_{\boldsymbol{\theta}_q, \boldsymbol{\theta}} \; \mathcal{L}(\boldsymbol{\theta}_q, \boldsymbol{\theta}, \boldsymbol{y}).
    \label{opt1}
\end{equation}

We remark that \eqref{opt1} can be conceived as two sequential optimization problems with respect to $\boldsymbol{\theta}_q$ and $\boldsymbol{\theta}$. For the former, maximizing  $\mathcal{L}$ is equivalent to minimizing the KL divergence (as an aforementioned goal) since $\log p(\boldsymbol{y}|\boldsymbol{\theta})$ in \eqref{logllh} is not a function of $\boldsymbol{\theta}_q$. For the latter, since the KL divergence has been minimized, the ELBO gives a good approximation of likelihood $\log p(\boldsymbol{y}|\boldsymbol{\theta})$. Therefore, the solution of \eqref{opt1} simultaneously provides the optimal $\boldsymbol{\theta}_q$ leading to a good approximation of $p(\boldsymbol{f}, \boldsymbol{g}|\boldsymbol{y})$ by $q(\boldsymbol{f}, \boldsymbol{g})$, as well as the optimal $\boldsymbol{\theta}$ leading to a maximized likelihood.

However, \eqref{opt1} with respect to $\boldsymbol{\theta}_q$ is still a prohibitively expensive optimization with dimensions $2n + 2n(2n+1)/2$ (i.e., number of unique elements in the mean and co-variance matrix of $q(\boldsymbol{f}, \boldsymbol{g})$). To reduce the number of dimensions, a key procedure employed in VHGPR is to assume that the function $q(\boldsymbol{f}, \boldsymbol{g})$ is separable in $\boldsymbol{f}$ and $\boldsymbol{g}$, i.e., $q(\boldsymbol{f},\boldsymbol{g}) = q_f(\boldsymbol{f})q_g(\boldsymbol{g})$. This brings two benefits: First, one can show that the maximized solution of $\mathcal{L}$ involves a relation between $q_f(\boldsymbol{f})$ and $q_g(\boldsymbol{g})$, i.e., $q_f(\boldsymbol{f})$ can be represented as a function of $q_g(\boldsymbol{g})$ so that the parameters in $\boldsymbol{\theta}_q$ is reduced to the mean $\boldsymbol{\mu}$ and covariance $\Sigma$ of $q_g(\boldsymbol{g})$ with dimension $n + n(n+1)/2$ (see (10.6) in \citep{bishop2006pattern} for proof). Second, the stationary point of $\mathcal{L}$ with respect to $\boldsymbol{\mu}$ and $\Sigma$ (by making $\partial \mathcal{L}/ \partial \boldsymbol{\mu} = \boldsymbol{0}$ and $\partial \mathcal{L}/ \partial \Sigma = \boldsymbol{0}$ ) leads to an analytical form
\begin{align}
    \boldsymbol{\mu} & = {\rm{K_g}} (\mathbb{X},\mathbb{X}) (\Lambda- \frac{1}{2} {\rm{I}}) \boldsymbol{1} + \mu_0 \boldsymbol{1}, \label{reducedmu}\\
    \Sigma& = ({\rm{K_g}}(\mathbb{X},\mathbb{X})^{-1} +  \Lambda)^{-1} \label{reducedsigma},
\end{align}
with ${\rm{K}}_g(\mathbb{X},\mathbb{X})_{ij}=k_g(\boldsymbol{x}^i,\boldsymbol{x}^j)$,  $\Lambda$ being a diagonal matrix involving $n$ unknown parameters and $\boldsymbol{1}$ being a vector with all elements 1. Therefore the optimization \eqref{opt1} is finally reduced to 
\begin{equation}
    \Lambda^*, \boldsymbol{\theta}^* = {\rm{argmax}}_{\Lambda, \boldsymbol{\theta}} \; \mathcal{L}(\boldsymbol{\mu}(\Lambda), \Sigma(\Lambda), \boldsymbol{\theta}, \boldsymbol{y}), \label{opt3}
\end{equation}
with only $n$ parameters in $\boldsymbol{\theta}_q$ (or $\Lambda$). This can be solved by gradient-based method, with the major computational cost in computing $n \times n$ matrix inversions in $\mathcal{L}$. The computational cost of each iteration in  \eqref{opt3} is only approximately twice as that in SGPR.

With $q_f(\boldsymbol{f})$, $q_g(\boldsymbol{g})$ (computed from $\Lambda^*$), and $\boldsymbol{\theta}$ available, the posterior predictions for $f$ and $g$ in \eqref{pred} are:
\begin{align}
    p(f(\boldsymbol{x}),g(\boldsymbol{x})|\mathcal{D}) & = \iint p(f(\boldsymbol{x}),g(\boldsymbol{x})|\boldsymbol{f},\boldsymbol{g}) q_f(\boldsymbol{f})q_g(\boldsymbol{g}) d \boldsymbol{f} d \boldsymbol{g}
\nonumber\\
    & = \iint p(f(\boldsymbol{x})|\boldsymbol{f}) p(g(\boldsymbol{x})|\boldsymbol{g}) q_f(\boldsymbol{f})q_g(\boldsymbol{g}) d \boldsymbol{f} d \boldsymbol{g} \nonumber \\ & \quad \quad({\rm{independence \; of}} \; f(\boldsymbol{x}) \; {\rm{and}} \; g(\boldsymbol{x})) 
\nonumber\\
    & = \int p(f(\boldsymbol{x})|\boldsymbol{f}) q_f(\boldsymbol{f}) d \boldsymbol{f} \int p(g(\boldsymbol{x})|\boldsymbol{g}) q_g(\boldsymbol{g})  d \boldsymbol{g}
\nonumber\\
    & = p(f(\boldsymbol{x})|\mathcal{D}) p(g(\boldsymbol{x})|\mathcal{D}),
\label{predfg} 
\end{align}
where: 
\begin{align}
    p(f(\boldsymbol{x})|\mathcal{D}) & =  \mathcal{N}(\mu_f(\boldsymbol{x}), {\rm{cov}}_f(\boldsymbol{x},\boldsymbol{x}')), 
\label{predf}\\ %1
    p(g(\boldsymbol{x})|\mathcal{D}) & =  \mathcal{N}(\mu_g(\boldsymbol{x}), {\rm{cov}}_g(\boldsymbol{x},\boldsymbol{x}')), 
\label{predg} \\ %2
    \mu_f(\boldsymbol{x}) & =  k_f(\boldsymbol{x},\mathbb{X})^T({\rm{K}}_f(\mathbb{X},\mathbb{X}) + Z)^{-1} \boldsymbol{y},    
\label{meanpredf} \\ %3
    {\rm{cov}}_f(\boldsymbol{x},\boldsymbol{x}')& =  k_f(\boldsymbol{x},\boldsymbol{x}') - k_f(\boldsymbol{x},\mathbb{X})^T ({\rm{K}}_f(\mathbb{X},\mathbb{X}) + Z)^{-1} k_f(\boldsymbol{x}',\mathbb{X}),
\label{covpredf}\\ %4
    \mu_g(\boldsymbol{x})& =   k_g(\boldsymbol{x},\mathbb{X})^T(\Lambda - \frac{1}{2}{\rm{I}})\boldsymbol{1} + \mu_0, 
\label{meanpredg} \\ %5
    {\rm{cov}}_g(\boldsymbol{x},\boldsymbol{x}') & = k_g(\boldsymbol{x},\boldsymbol{x}') - k_g(\boldsymbol{x},\mathbb{X})^T ({\rm{K}}_g(\mathbb{X},\mathbb{X}) + \Lambda^{-1})^{-1} k_g(\boldsymbol{x}',\mathbb{X}), 
\label{covpredg} %6
\end{align}
with $Z_{ii} = e ^{\mu_i-\Sigma_{ii} /2}$ being a diagonal matrix and ${\rm{K}}_f(\mathbb{X},\mathbb{X})_{ij}=k_f(\boldsymbol{x}^i,\boldsymbol{x}^j)$.

%%%%%%%%%%%%%%%%%%%%%%%%%%%%%%%%%%%%%%%%%%%%%%%%%%%%%%
\section{The upper bound of the estimation variance}
Here we show the construction of an upper bound of the estimation variance (For convenience we use $\hat{p}(\boldsymbol{x})$ to represent $\mathbb{P}(S(\boldsymbol{x},\omega| f(\boldsymbol{x}), g(\boldsymbol{x}))> \delta)$):
\begin{align}
    &{\rm{var}}_{f,g} [\mathbb{P}(S(X,\omega| f(\boldsymbol{x}), g(\boldsymbol{x}))> \delta)] 
\nonumber\\ %1
    =  & \; {\rm{var}}_{f,g}   \Big[ \int \hat{p}(\boldsymbol{x})    p_X(\boldsymbol{x}) d \boldsymbol{x} \Big] 
\nonumber  \\ %2
    = & \; \mathbb{E}_{f,g} \Big[\Big(\int \hat{p}(\boldsymbol{x}) p_X(\boldsymbol{x}) d \boldsymbol{x} \Big)^2 \Big] -  \Big(\mathbb{E}_{f,g} \Big[\int \hat{p}(\boldsymbol{x}) p_X(\boldsymbol{x}) d \boldsymbol{x}\Big]\Big)^2 
\nonumber \\ %3
    = & \; \mathbb{E}_{f,g} \Big[\int \hat{p}(\boldsymbol{x}) p_X(\boldsymbol{x}) d \boldsymbol{x} \int \hat{p}(\boldsymbol{x}') p_X(\boldsymbol{x}') d \boldsymbol{x}'\Big] 
\nonumber \\ %4
    & -  \Big(\mathbb{E}_{f,g} \Big[\int \hat{p}(\boldsymbol{x}) p_X(\boldsymbol{x}) d \boldsymbol{x}\Big]\Big) \Big(\mathbb{E}_{f,g} \Big[\int \hat{p}(\boldsymbol{x}') p_X(\boldsymbol{x}') d \boldsymbol{x}'\Big]\Big)
\nonumber \\ %5
    = & \iint \mathbb{E}_{f,g} \Big[\hat{p}(\boldsymbol{x})\hat{p}(\boldsymbol{x}')\Big] p_X(\boldsymbol{x}) p_X(\boldsymbol{x}') d \boldsymbol{x} d \boldsymbol{x}' 
\nonumber\\ %6
    & -  \iint \mathbb{E}_{f,g} \Big[\hat{p}(\boldsymbol{x})\Big] \mathbb{E}_{f,g} \Big[ \hat{p}(\boldsymbol{x}')\Big] p_X(\boldsymbol{x}) p_X(\boldsymbol{x}') d \boldsymbol{x} d \boldsymbol{x}' 
\nonumber\\ %7
    = & \iint {\rm{cov}}_{f,g} \big[\hat{p}(\boldsymbol{x}), \hat{p}(\boldsymbol{x}')\big] p_X(\boldsymbol{x}) p_X(\boldsymbol{x}') d \boldsymbol{x} d\boldsymbol{x}' 
\nonumber \\ %8
    = & \int {\rm{std}}_{f,g} \big[\hat{p}(\boldsymbol{x})\big]  \Big( \int {\rm{std}}_{f,g} \big[\hat{p}(\boldsymbol{x}')\big] \rho  \big[ \hat{p}(\boldsymbol{x}),\hat{p}(\boldsymbol{x}')\big] p_X(\boldsymbol{x}') d\boldsymbol{x}' \Big) p_X(\boldsymbol{x})  d \boldsymbol{x}
\nonumber \\ %9
    \leq & \; 0.5 \int {\rm{std}}_{f,g} \big[\hat{p}(\boldsymbol{x})\big] p_X(\boldsymbol{x}) d \boldsymbol{x} 
\label{bound} %10
\end{align}
where $\rho[\cdot,\cdot]$ denotes the correlation coefficient. The last inequality comes from  ${\rm{std}} [\hat{p}_{\omega}(\boldsymbol{x}')] \leq 0.5$  and  $\rho [\hat{p}_{\omega}(\boldsymbol{x}),\hat{p}_{\omega}(\boldsymbol{x}')] \leq 1$. The equality in \eqref{bound} holds when %${\rm{std}} [\hat{p}_{\omega}(\boldsymbol{x})] = 0.5$ 
 $\hat{p}_{\omega}$ degenerates to a Bernoulli random variable with equal probability 0.5 at both $\hat{p}_{\omega}= 0$ and $\hat{p}_{\omega}= 1$ and  $\rho [\hat{p}_{\omega}(\boldsymbol{x}),\hat{p}_{\omega}(\boldsymbol{x}')] = 1$. 

%\section{The Spherical Cubature Integration}
%The spherical cubature integration is special case of unscented Kalman filter originally proposed to solve filtering problems of nonlinear functions \citep{wan2000unscented}. The Cubature filters is third-order accuracy with $2d$ quadrature points and weights where $d$ is the dimension of the integration. 

\section{Spherical cubature integration in equation (12)-(14)}
For given $\boldsymbol{x}$, both equations \eqref{mean} and \eqref{var} can be considered as two-dimensional ($d=2$) Gaussian weighted integrals. Let $\boldsymbol{u}$ denote $\{f(\boldsymbol{x}), g(\boldsymbol{x})\}$ and $h(\boldsymbol{u})$ denote $\mathbb{P} (S(\boldsymbol{x}, \omega |\boldsymbol{u}) > \delta)$ or $(\mathbb{P} (S(\boldsymbol{x}, \omega |\boldsymbol{u}) > \delta)-m)^2$. Both \eqref{mean} and \eqref{var} can be rewritten in the following form: 
\begin{equation}
    \int h(\boldsymbol{u}) \mathcal{N}(\boldsymbol{u};\boldsymbol{\mu}, \Sigma) d\boldsymbol{u},
    \label{C1}
\end{equation}
where
\begin{align}
    \boldsymbol{\mu} & = [\mu_f(\boldsymbol{x}) \; \; \mu_g(\boldsymbol{x})]^T, 
\\
    \Sigma           & = 
\left[ \begin{matrix}  
{\rm{cov}}_f(\boldsymbol{x}) & 0 \\
 0                           & {\rm{cov}}_g(\boldsymbol{x}) \\
\end{matrix} \right].
\end{align}
We further define a standard Gaussian random vector $\tilde{\boldsymbol{u}}=\sqrt{\Sigma}^{-1}({\boldsymbol{u}}-\boldsymbol{\mu})$, where 
\begin{equation}
    \sqrt{\Sigma}            = 
\left[ \begin{matrix}  
{\sqrt{\rm{cov}}_f(\boldsymbol{x})} & 0 \\
 0                           & {\sqrt{\rm{cov}}_g(\boldsymbol{x})} \\
\end{matrix} \right],
\end{equation}
or more general, the Cholesky decomposition of $\Sigma$. Then \eqref{C1} can be transformed to 
\begin{equation}
   \int h(\boldsymbol{\mu} +\sqrt{\Sigma}\tilde{\boldsymbol{u}})\mathcal{N}(\tilde{\boldsymbol{u}}; \boldsymbol{0}, {\rm{I}}) d\tilde{\boldsymbol{u}} 
    = \int \hat{h}(\tilde{\boldsymbol{u}})\mathcal{N}(\tilde{\boldsymbol{u}}; \boldsymbol{0}, {\rm{I}}) d\tilde{\boldsymbol{u}}, 
\label{scaleI}
\end{equation}
where we have defined $h(\boldsymbol{\mu} +\sqrt{\Sigma}\tilde{\boldsymbol{u}})=\hat{h}(\tilde{\boldsymbol{u}})$ for simplicity. 

The spherical cubature integration aims to approximate \eqref{scaleI} with $2d$ points \citep{sarkka2013bayesian}:
\begin{equation}
    \int \hat{h}(\tilde{\boldsymbol{u}})\mathcal{N}(\tilde{\boldsymbol{u}}; \boldsymbol{0}, {\rm{I}}) d\tilde{\boldsymbol{u}} \approx w \sum_{i=1}^{2d = 4} \hat{h}(c\tilde{\boldsymbol{u}}^{(i)}),
\end{equation}
where $\tilde{\boldsymbol{u}}^{(i)} \in \{\{1,0\}, \{0,1\}, \{-1,0\}, \{0,-1\}\}$, and $w$ and $c$ are coefficients determined by satisfying the following conditions for third order accuracy: 
\begin{align}
    \int \mathcal{N}(\tilde{\boldsymbol{u}}; \boldsymbol{0}, {\rm{I}}) d \tilde{\boldsymbol{u}} & = w \sum_{i=1}^4 1,
\\
    \int \tilde{u}_j\mathcal{N}(\tilde{\boldsymbol{u}}; \boldsymbol{0}, {\rm{I}}) d \tilde{\boldsymbol{u}} & = w \sum_{i=1}^4 (c \tilde{u}_j^{(i)}),
\\
    \int \tilde{u}_j^2\mathcal{N}(\tilde{\boldsymbol{u}}; \boldsymbol{0}, {\rm{I}}) d \tilde{\boldsymbol{u}} & = w \sum_{i=1}^4 (c \tilde{u}_j^{(i)})^2, \quad j = 1,2 .
\end{align}
According to \citep{sarkka2013bayesian}, this yields $w = 1/4$ and $c = \sqrt{2}$.

Finally, the values of $\tilde{\boldsymbol{u}}^{(i)}$ can be transformed back to ${\boldsymbol{u}}^{(i)}= \boldsymbol{\mu} +\sqrt{\Sigma}\tilde{\boldsymbol{u}}^{(i)}$, corresponding to \eqref{u} in Sec. 2.3.


\end{document}